\documentclass[a4paper]{jpconf}
\usepackage{graphicx}
\usepackage{amsmath}

\def\blue{\color{blue}}

\def\black{\color{black}}
\def\sech{\rm sech}

\def\tanh{\rm tanh}
\def\half{{1\over 2}}

\def\Z{{\mathchoice {\hbox{$\sf\textstyle Z\kern-0.4em Z$}}
{\hbox{$\sf\textstyle Z\kern-0.4em Z$}}
{\hbox{$\sf\scriptstyle Z\kern-0.3em Z$}}
{\hbox{$\sf\scriptscriptstyle Z\kern-0.2em Z$}}}}

\def\square{\kern1pt\vbox{\hrule height 1.2pt\hbox{\vrule width 1.2pt
   \hskip 3pt\vbox{\vskip 6pt}\hskip 3pt\vrule width 0.6pt}
   \hrule height 0.6pt}\kern1pt}
      \def\boxop{{\raise-.25ex\hbox{\square}}}

\def\e{\,{\rm e}}

\def\be{\begin{equation}}
\def\ee{\end{equation}\noindent}
\def\bear{\begin{eqnarray}}
\def\ear{\end{eqnarray}\noindent}
\def\bec{\blue\begin{equation}}
\def\eec{\end{equation}\black\noindent}
\def\bearc{\blue\begin{eqnarray}}
\def\earc{\end{eqnarray}\black\noindent}
\def\benn{\begin{enumerate}}
\def\enn{\end{enumerate}}

\def\slash#1{#1\!\!\!\raise.15ex\hbox {/}}
\newcommand{\slD}{\,\raise.15ex\hbox{$/$}\kern-.27em\hbox{$\!\!\!D$}}
\newcommand{\slpartial}{\raise.15ex\hbox{$/$}\kern-.57em\hbox{$\partial$}}

\newcommand{\nc}{\newcommand}
\nc{\spa}[3]{\left\langle#1\,#3\right\rangle}
\nc{\spb}[3]{\left[#1\,#3\right]}
\nc{\ksl}{\not{\hbox{\kern-2.3pt $k$}}}
\nc{\hf}{\textstyle{1\over2}}
\nc{\pol}{\varepsilon}
\nc{\tq}{{\tilde q}}
\nc{\esl}{\not{\hbox{\kern-2.3pt $\pol$}}}

\def\4piTD{{(4\pi T)}^{-{D\over 2}}}
\def\4piT4{{(4\pi T)}^{-2}}

\def\Tintm4{{\dps\int_{0}^{\infty}}{dT\over T}\,e^{-m^2T}
    {(4\pi T)}^{-2}}
\def\Tintm{{\dps\int_{0}^{\infty}}{dT\over T}\,e^{-m^2T}}

\newcommand{\slG}{{{\dot G}\!\!\!\! \raise.15ex\hbox {/}}}

\def\GBd12{{\dot G}_{B12}}

\def\dps{\displaystyle}

\begin{document}
\title{Generalized Gelfand-Dikii equation for fermionic Schwinger pair production}

\author{N Ahmadiniaz$^1$, S P Kim$^{2,3}$ and C Schubert$^{4,5}$}

\address{$^1$Helmholtz-Zentrum Dresden-Rossendorf, Bautzner Landstra\ss e 400, 01328 Dresden, Germany}

\address{$^2$Department of Physics, Kunsan National University, Kunsan 54150, Korea}

\address{$^3$Center for Relativistic Laser Science, Institute for Basic Science,
Gwangju 61005, Korea}

\address{$^4$Instituto de F\'{i}sica y Matem\'{a}ticas Universidad Michoacana de San Nicol\'{a}s de Hidalgo
Edificio C-3, Apdo. Postal 2-82 C.P. 58040, Morelia, Michoac\'{a}n, Mexico}

\address{$^5$Centro Internacional de Ciencias A.C. Campus UNAM-UAEM Cuernavaca Mor. Mexico C.P. 62100}

\ead{n.ahmadiniaz@hzdr.de,sangkim@kunsan.ac.kr,christianschubert137@gmail.com}

\begin{abstract}
Schwinger pair creation in a purely time-dependent electric field can be reduced to an effective quantum mechanical problem using a variety of formalisms. Here we develop an approach based on the Gelfand-Dikii equation for scalar QED, and on a generalization of that equation for spinor QED. We discuss a number of solvable special cases from this point of view. In previous work, two of the authors had shown for the scalar case how to use the well-known solitonic solutions of the KdV equation to construct Pöschl-Teller like electric fields that do not pair create at some fixed but arbitrary momentum. Here, we present numerical evidence that this construction can be adapted to the fermionic case by a mere change of parameters.
\end{abstract}

\centerline{Published in {\sl J. Phys.: Conf. Ser. 2249 012020}}

\section{Introduction}

The case of a purely time-dependent field is, despite of the idealized character of such a field, a 
popular testing ground for approaches to Schwinger pair creation in strong-field QED. This is
because it can effectively be reduced to a quantum-mechanical time evolution problem, and thus is
relatively amenable to exact analytical treatments. Moreover, this time evolution can be formulated
in many equivalent ways (see, e.g., \cite{dabdun}). Here we focus on an approach based on the
Gelfand-Dikii equation, which offers an interesting link to the Korteweg-de-Vries (`KdV') equation. In previous work by two
of the authors \cite{84} this connection was used for the construction of electric fields in scalar QED that are related to the well-known solitonic solutions of that equation, and have the property that they do {\it not} pair create at some fixed reference momentum. Here we extend this approach to the spinor QED
case. We find an evolution equation describing the pair creation of spin half particles that can be
considered as a natural generalization of the Gelfand-Dikii equation, and we present numerical
evidence that also the concept of solitonic electric fields possesses some extension to the
spinor QED case.

\section{Basic set-up}

Let us consider a purely time-dependent electric field $ {\bf E}(t)$ pointing into a fixed direction, and its vector potential in 
temporal gauge $ A_0 = 0, \dot {\bf A}(t) = - {\bf E}(t)$. 
We further assume that the field is localized in time, $ E(\infty) = E(-\infty)=0$, with a finite integral
$ \int_{-\infty}^{\infty} dt\, E(t) = A(-\infty)-A(\infty)$.

\section{Review of the scalar QED case}

\subsection{The mode equation}

The Hamiltonian of a charged field with mass $m$ and charge $q$ in a purely
time-dependent electric field can be decomposed into an
infinite number of harmonic oscillators with time-dependent frequencies,
\begin{eqnarray}
H(t) = \sum_{\bf k} \pi_{\bf k}^* \pi_{\bf k} + \omega_{\bf k}^2 (t) \phi_{\bf k}^* \phi_{\bf k}, 
\end{eqnarray}
where 
\begin{eqnarray}
\omega_{\bf k}^2 (t) = \Pi^2 + \mu^2, \quad \Pi = k_{\parallel} - qA_{\parallel}, \quad \mu^2 = m^2 + {\bf k}_{\perp}^2
\end{eqnarray}
and $ k_{\parallel}, k_\perp$ denote the components of the canonical three-momentum along the field resp. perpendicular to it. 
Since we assume that the field integral is finite, we can define the limits
\begin{eqnarray}
\omega_{\bf k}^{\pm} \equiv \lim_{t\to\pm \infty} \omega_{\bf k}(t)
\, .
\end{eqnarray}
The simplicity of the purely time-dependent electric field case is due to the fact that the
individual modes $ \phi_{\bf k}$ obey a time-dependent oscillator equation (“ mode equation”)
\begin{eqnarray}
\ddot{\phi}_{\bf k} (t) + \omega^2_{\bf k} (t) \phi_{\bf k} (t) = 0 \, .
\label{meqscal}
\end{eqnarray}
Canonical quantization of the scalar field leads to the  Wronskian constraint
\begin{eqnarray}
{\rm Wr} [\phi_{\bf k}, \phi^*_{\bf k}] \equiv   \phi_{\bf k}\dot\phi^*_{\bf k} - \dot\phi_{\bf k}\phi^*_{\bf k}=  i
\, .
\label{wronskian}
\end{eqnarray}
Using  Bogolyubov theory, from a solution of the mode equation with index $\bf k$ one can get the
 density of created pairs $ {\cal N}_{\bf k} (t)$
\begin{eqnarray}
{\cal N}_{\bf k}(t) = \frac{|\dot \phi_{\bf k} (t)|^2 + \omega_{\bf k}^2(t) |\phi_{\bf k} (t) |^2}
{2\omega_{\bf k}(t)} - \frac{1}{2}
\, .
\label{relNphi}
\end{eqnarray}
Usually one assumes that initially there are no particles present, $ {\rm lim}_{t\to -\infty}{\cal N}_{\bf k}(t)=0$.
The relevant solution of the mode equation will then obey (up to an irrelevant phase factor)
\begin{eqnarray}
\phi_{\bf k}(t)\, \stackrel{t\to -\infty}{\longrightarrow}\, \frac{1}{\sqrt{2\omega_i}} \,\e^{-i\omega_i t} 
\, .
\label{condin}
\end{eqnarray}
No direct physical meaning can be ascribed to ${\cal N}_{\bf k}(t)$ at intermediate times, 
since its definition is inherently ambiguous on account of its dependence on the choice of an instantaneous adiabatic basis \cite{dabdun}. 

\subsection{The quantum Vlasov equation}

Alternatively, one can use the  Quantum Vlasov Equation 
\cite{kescm91,kescm92,sbrspt,healgi}.
This is an evolution equation at fixed $\bf k$ for the density of pairs $ {\cal N}_{\bf k}(t)$:
\begin{eqnarray}
\dot{\cal N}_{\bf k} (t)
&=& \frac{\dot\omega_{\bf k} (t)}{2\omega_{\bf k} (t)} \int_{t_0}^{t} dt'  \frac{\dot\omega_{\bf k} (t')}{\omega_{\bf k} (t')}
(1+2 {\cal N}_{\bf k} (t') )  \cos \Bigl[ 2 \int_{t'}^{t} dt'' \omega_{\bf k} (t'')  \Bigr]
\, .
\end{eqnarray}
Here $ t_0$ is the initial time, usually $ -\infty$, and
\begin{eqnarray}
\omega_{\bf k}^2 (t) &=& (k_{\parallel} - qA_{\parallel} (t))^2 + {\bf k}_{\perp}^2 + m^2
\, .
\end{eqnarray}
$ {\cal N}_{\bf k}(t)$ is zero at $ t = -\infty$, and for $ t\to\infty$ turns into the density of created pairs
with fixed momentum $\bf k$.

\subsection{The time-like Sauter field}

\begin{figure}[h]
\begin{center}
\includegraphics[scale=.5]{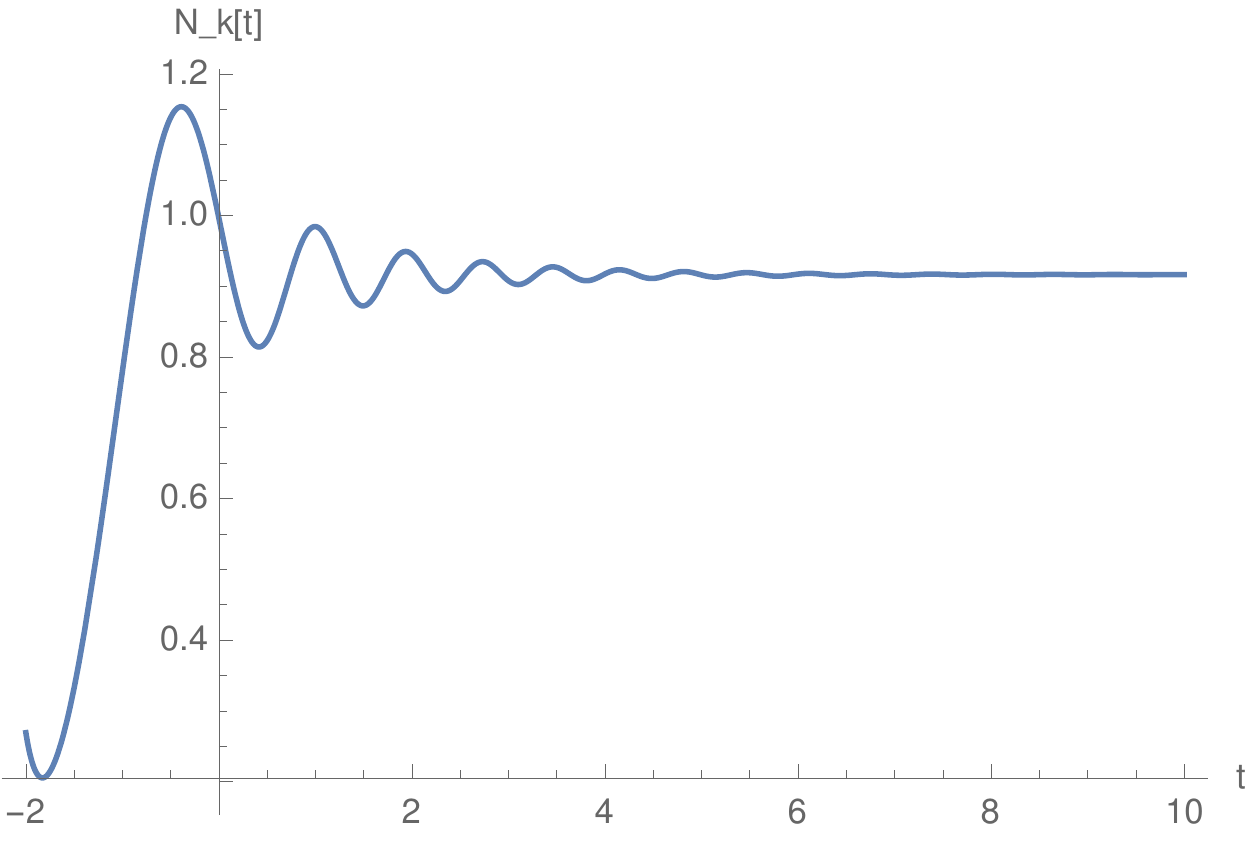}
\end{center}
\caption{Plot of the pair-creation rate ${\cal N}_{\bf k}(t)$ for the Sauter field with $E_0=1$, $\tau=3$, $k_{\parallel} = \omega_0=1$.}
\label{fig-Nsauter}
\end{figure}

As an example, let us consider the time-like Sauter field, defined by
\begin{eqnarray}
E(t)=-\dot{A} = -E_0 \,{\rm sech}^2(t/\tau) 
\end{eqnarray}
%
%
%
which we can realize by 
\begin{eqnarray}
A^{\mu} = (0,0,0,E_0 \tau (1+\tanh (t/\tau) )) \, .
\end{eqnarray}
The exact solution of the mode equation for this field, obeying the initial conditions,
is given by
\begin{eqnarray}
\phi_{\bf k}(t) &=& \frac{1}{\sqrt{2\omega_i \e^{\pi\omega_i\tau}}}z^{-\frac{i}{2}\omega_i\tau}
(1-z)^{\frac{1}{2} + i \lambda^{\rm sc}}
\phantom{}_2F_1(\alpha^{\rm sc},\beta^{\rm sc};\gamma;z)
\end{eqnarray}
where $z\equiv -\e^{\frac{2t}{\tau}}$, and
\begin{eqnarray}
\lambda^{\rm sc} &\equiv& \sqrt{(eE_0\tau^2)^2-\frac{1}{4}} \, ,
\nonumber\\
\alpha^{\rm sc} &\equiv & \frac{1}{2} - \frac{i}{2} \Bigl\lbrack \tau(\omega_i - \omega_f) - 2\lambda^{\rm sc} \Bigr\rbrack \, ,
\nonumber\\
\beta^{\rm sc} &\equiv & \frac{1}{2} - \frac{i}{2} \Bigl\lbrack \tau(\omega_i + \omega_f) - 2\lambda^{\rm sc} \Bigr\rbrack  \, ,
\nonumber\\
\gamma &\equiv& 1- i\tau\omega_i  \, ,
\nonumber\\
\omega_i&=&\omega_0 \equiv \sqrt{{\bf k}^2+m^2} \, ,\nonumber\\
\omega_f&=& \sqrt{(k_{\parallel}-2qE_0\tau)^2+\mu^2} \, .\nonumber\\
\end{eqnarray}
Using the solution of the mode equation in (\ref{relNphi}), we plot
${\cal N}_{\bf k}(t)$ for $ E_0=1$, $ \tau=3$, and $ k_{\parallel} = \omega_0=1$ (Fig. \ref{fig-Nsauter})

\vspace{20pt}

Only its asymptotic value for $ t\to\infty$ is physically meaningful.

\subsection{The Gelfand-Dikii equation}

To obtain $ {\cal N}_{\bf k}(t)$, it is sufficient to know $ |\phi_{\bf k}(t)|$. This suggests that it might be more
economical to work with
$ G_{\bf k} (t) \equiv |\phi_{\bf k}(t)|^2$, which evolves by a linear third-order differential equation,
\begin{eqnarray}
\dddot{G}_{\bf k} (t) + 4 \omega^2_{\bf k} (t) \dot{G}_{\bf k} (t) + 2 \bigl(\omega_{\bf k}^2 (t) \bigr)^{\cdot} G_{\bf k} (t) = 0
\label{geldikeq}
\end{eqnarray}
with the initial condition
\begin{eqnarray}
{\rm lim}_{t\to -\infty} G_{\bf k}(t) = \frac{1}{2\omega_i}
\, .
\nonumber
\end{eqnarray}
This is actually the  Gelfand-Dikii equation \cite{geldik}.
In terms of $ G_{\bf k}$, $ {\cal N}_{\bf k}(t)$ is given by
\begin{eqnarray}
{\cal N}_{\bf k}(t) = \frac{\ddot G_{\bf k}}{4\omega(t)} +\omega(t) G_{\bf k} - \frac{1}{2}
\, .
\end{eqnarray}
For example, for the Sauter field one has 
\begin{eqnarray}
G_{\bf k} &=& \frac{1-z} {2\omega_0 \e^{\pi\omega_0\tau}}
\Bigl\vert \phantom{}_2F_1 (\alpha^{\rm sc},\beta^{\rm sc};\gamma;z) \Bigr\vert^2
\, .
\end{eqnarray}

\subsection{Solitonic electric fields}

\begin{figure}[htp]
\vspace{-2pt}
\begin{center}
\includegraphics[scale=.5]{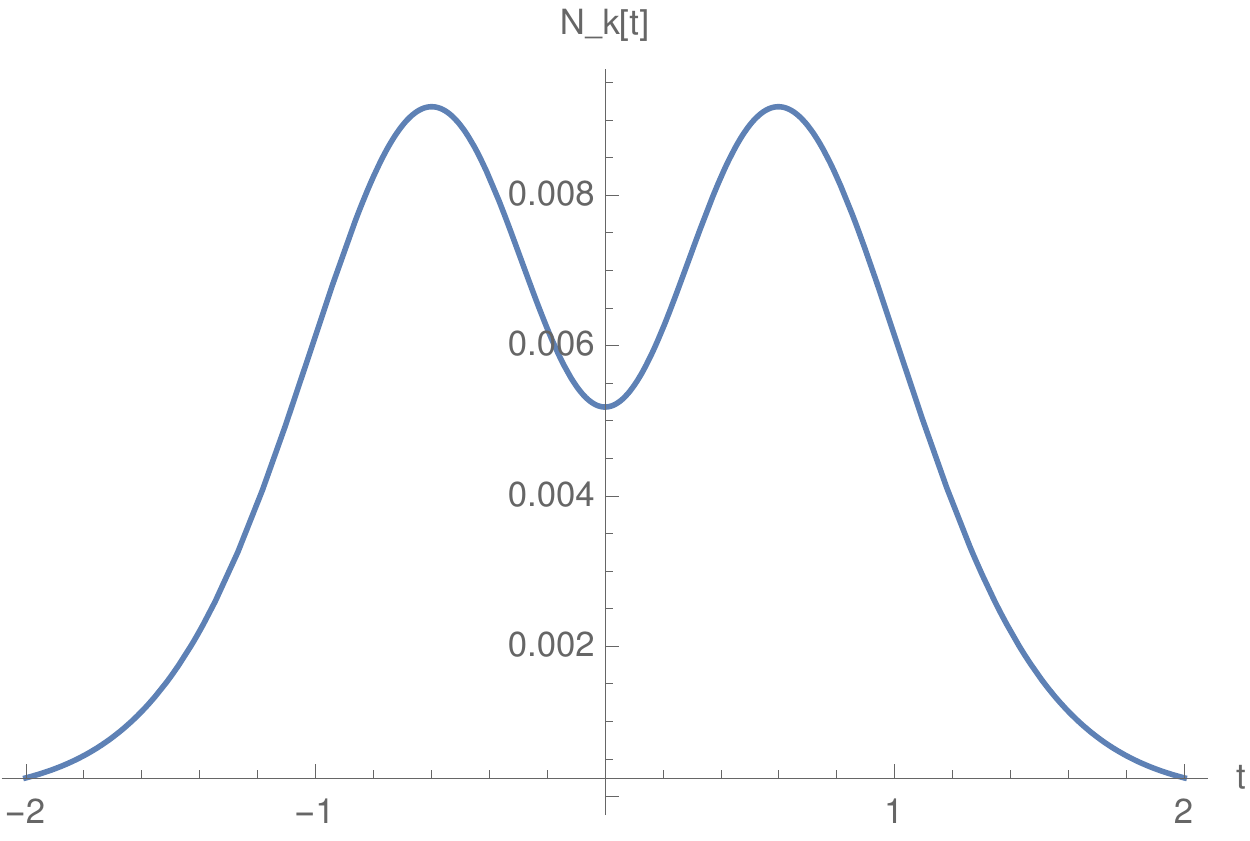}
\end{center}
\caption{${\cal N}_{\bf\tilde  k}(t)$ for the $p=1$ soliton (with $\omega_0=1.1$).} 
\label{fig-soliton}
\end{figure}

The Gelfand-Dikii equation relates to the resolvent of the mode equation, and through it to the (generalized) KdV equation \cite{geldik}. 
In \cite{84,101} the following infinite family of vector potentials was constructed, related
to the solitonic solutions of the KdV equation:
\begin{eqnarray}
qA_{\parallel(p)}(t) \equiv  \tilde k_{\parallel} - \sqrt{\tilde k_{\parallel}^2+ \frac{p(p+1)\omega_0^2}{\cosh^2(\omega_0 t)}}, \quad p=1,2,\ldots
\label{solitons}
\end{eqnarray}
where $ \tilde k_{\parallel}$ is a fixed but arbitrary reference momentum. 
The corresponding squared energies at $ k_{\parallel} = \tilde k_{\parallel}$ are of  P\"oschl-Teller type,
\begin{eqnarray}
\omega^2_{(p){\bf \tilde k}}(t) = \omega_0^2 \Bigl(1+ \frac{p(p+1)}{\cosh^2 (\omega_0 t)}\Bigr),
\quad p =1,2,\ldots 
\nonumber
\end{eqnarray}
All these ``solitonic'' electric fields do not produce pairs for $ k_{\parallel} = \tilde k_{\parallel}$
(although they will generally do so for other values of $ k_{\parallel}$). 
%
%
%
%
For these solitonic fields, the mode equation can be solved exactly:
\begin{eqnarray}
\phi_{(p)}(t)=\frac{e^{-i\omega_0t}}{\sqrt{2\omega_0}}\,\Big(1+e^{2\omega_0t}\Big)^{p+1}\,_2F_1\Big[1+p;-i+p+1;1-i;-e^{2\omega_0t}\Big]\, .
\end{eqnarray}
Let us show the resulting $ {\cal N}_{\tilde {\bf k}} (t)$ for the simplest case $ p=1$:
\begin{eqnarray}
{\cal N}_{\bf\tilde  k}(t) &=& \frac{4+\sech^4(\omega_0 t)(1+2\cosh(2\omega_0 t))}{8\sqrt{1+2\sech^2(\omega_0 t)}}- \half \, .
\end{eqnarray}
In Fig. \ref{fig-soliton} we plot this function for $\omega_0=1.1$.

Since $ {\cal N}_{\bf \tilde k}(t)$ vanishes for $ t\to\infty$
there is no pair creation at that particular momentum $ \bf k = \bf \tilde k$. 
The external field excites the vacuum, but no particles materialize.

\section{Generalization to spinor QED}

\subsection{The mode equation}

For spin $ \frac{1}{2}$ particles the mode equation becomes
\cite{basefr} 
\begin{eqnarray}
\ddot{\psi}_{\bf k}^{(\pm)} (t) + \bigl[\omega^2_{\bf k} (t) \mp iq \dot A_{\parallel}(t)\bigr]  \psi_{\bf k}^{(\pm)} (t) = 0 \, ,
\label{meqspin}
\end{eqnarray}
where the $(\pm)$ denotes the spin. 
In QFT, this comes from the spin-diagonalized square of the Dirac operator.
In \cite{fgks} it was shown that the constraint equation (\ref{wronskian}) generalizes to
\begin{eqnarray}
\omega_{\bf k}^2 \vert \psi_{\bf k}^{(\pm)}\vert^2 + \vert\dot\psi_{\bf k}^{(\pm)}\vert^2 \mp i\Pi
(\dot\psi_{\bf k}^{(\pm)} \psi_{\bf k}^{(\pm)\ast} - \psi_{\bf k}^{(\pm)}\dot\psi_{\bf k}^{(\pm)\ast}) = \mu^2
\label{wronskianspin}
\end{eqnarray}
where $\Pi=k_\parallel-qA_{\parallel}(t)$ and that ${\cal N}_{\bf k}(t)$ 
can, with appropriate initial conditions imposed on $ \psi_{\bf k}^{(\pm)}$, be written in terms of the mode solution as
\begin{eqnarray}
 {\cal N}_{\bf k}(t) &=& \frac{1}{2\omega_{\bf k} (\omega_{\bf k} - \Pi)} 
\biggl\lbrack
\omega_{\bf k}^2 |\psi_{\bf k}^{(\pm)}|^2 + |\dot \psi^{(\pm)}|^2 \mp i\omega_{\bf k}
(\dot\psi_{\bf k}^{(\pm)} \psi_{\bf k}^{(\pm)\ast} - \psi_{\bf k}^{(\pm)}\dot\psi^{(\pm)\ast}) 
\biggr\rbrack
\, .
\end{eqnarray}
Note that $ {\cal N}_{\bf k}(t)$ does not depend on the spin orientiation.
Combining the last two equations, we can also write
\begin{eqnarray}
{\cal N}_{\bf k}(t) &=& \frac{\mu^2}{2\Pi (\omega_{\bf k} - \Pi)} 
-
\frac{\omega_{\bf k}^2 |\psi_{\bf k}^{(\pm)}|^2 + |\dot \psi_{\bf k}^{(\pm)}|^2}
{2\omega_{\bf k}\Pi}
\, .
\end{eqnarray}

\subsection{Fermionic Quantum Vlasov Equation}

There is also a fermionic generalization of the Quantum Vlasov Equation \cite{sbrspt}
\begin{eqnarray}
\dot{\cal N}_{\bf k} (t)
&=& \frac{\mu q \dot A_{\parallel}(t)}{2 \omega_{\bf k}^2 (t)}
 \int_{-\infty}^{t} dt' \, 
 \frac{\mu q \dot A_{\parallel}(t')}{\omega_{\bf k}^2 (t')}
\bigl(1-2 {\cal N}_{\bf k} (t') \bigr)  \cos \Bigl\lbrack 2\int_{t'}^{t} dt'' \omega_{\bf k} (t'') \Bigr\rbrack  
\, .
\label{qvespin}
\end{eqnarray}

\subsection{Fermionic Gelfand-Dikii equation}

Defining $ G^\pm_{\bf k} \equiv \vert\psi_{\bf k}^{\pm}\vert^2$ like in the scalar case 
and combining (\ref{meqspin}), (\ref{wronskianspin}), one arrives
at the following generalization of the Gelfand-Dikii equation \eqref{geldikeq}
\begin{equation}
\dddot{G^\pm_{\bf k}}-\frac{E}{\Pi}\ddot{G^\pm_{\bf k}}+4\omega_{\bf k}^2\dot{G_{\bf k}}+\Bigl(2
\bigl(\omega_{\bf k}^2 \bigr)^{\cdot} 
-4\frac{E\omega_{\bf k}^2}{\Pi}\Bigr)G^\pm_{\bf k}=-2\mu^2\frac{E}{\Pi} \, .
\label{geldikeqspin}
\end{equation}
Differentiating the mode equation one can eliminate $ |\dot \psi^{(\pm)}|^2$ and rewrite $ {\cal N}_{\bf k}$ in terms of $ G_{\bf k}^\pm$:
\begin{eqnarray}
{\cal N}_{\bf k} = \frac{1}{2} + \frac{\omega_{\bf k}}{\Pi}\Bigl(\frac{1}{2}-G_{\bf k}^\pm\Bigr) - \frac{\ddot G_{\bf k}^\pm}{4\omega_{\bf k} \Pi}
\, .
\end{eqnarray}

\subsection{Fermions in the solitonic field}
\label{fermsol}

\begin{figure}[htp]
\vspace{-90pt}
\begin{center}
\includegraphics[scale=.4]{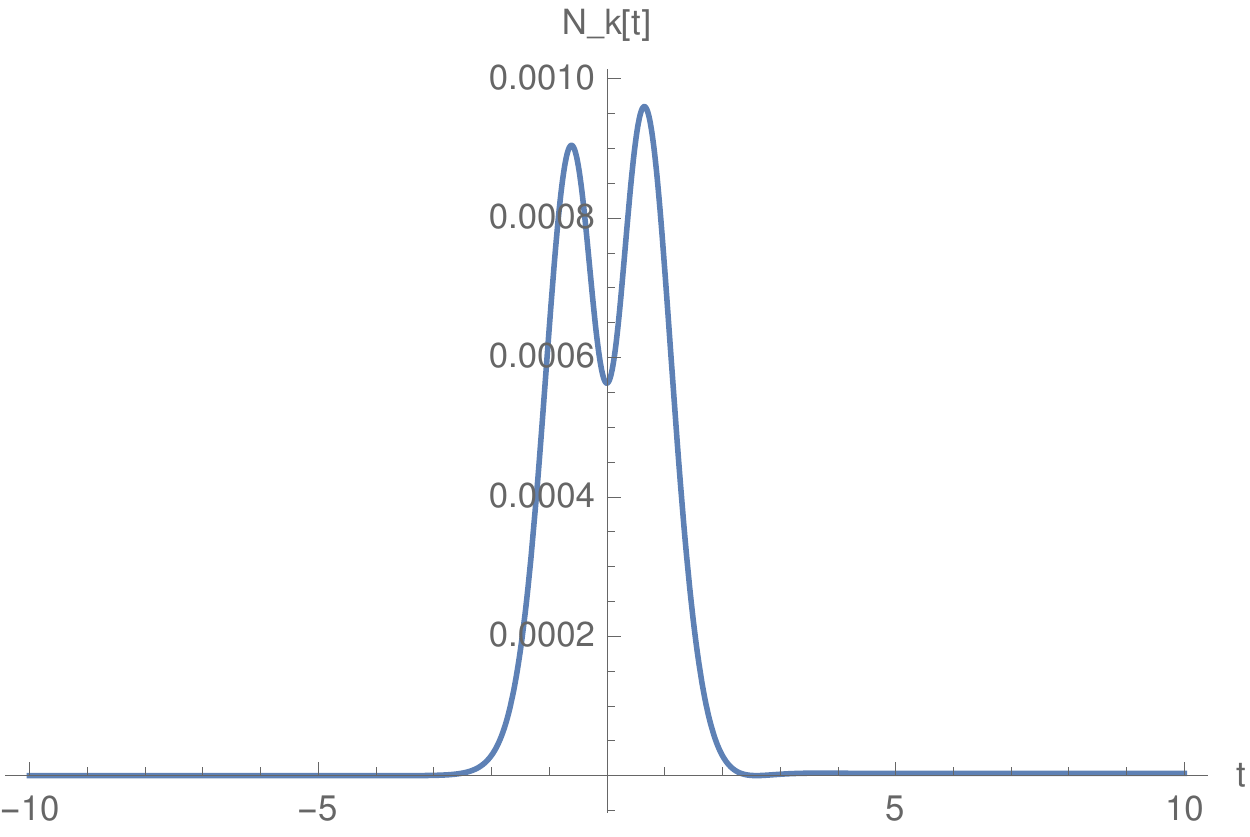}
\end{center}
\caption{Fermionic ${\cal N}_{\bf\tilde  k}(t)$ for the $p= \half \bigl(\pm \sqrt{7} -1\bigr)$ soliton.} 
\label{fig-m2}
\end{figure}

\begin{figure}[htp]
\begin{center}
\includegraphics[scale=.4]{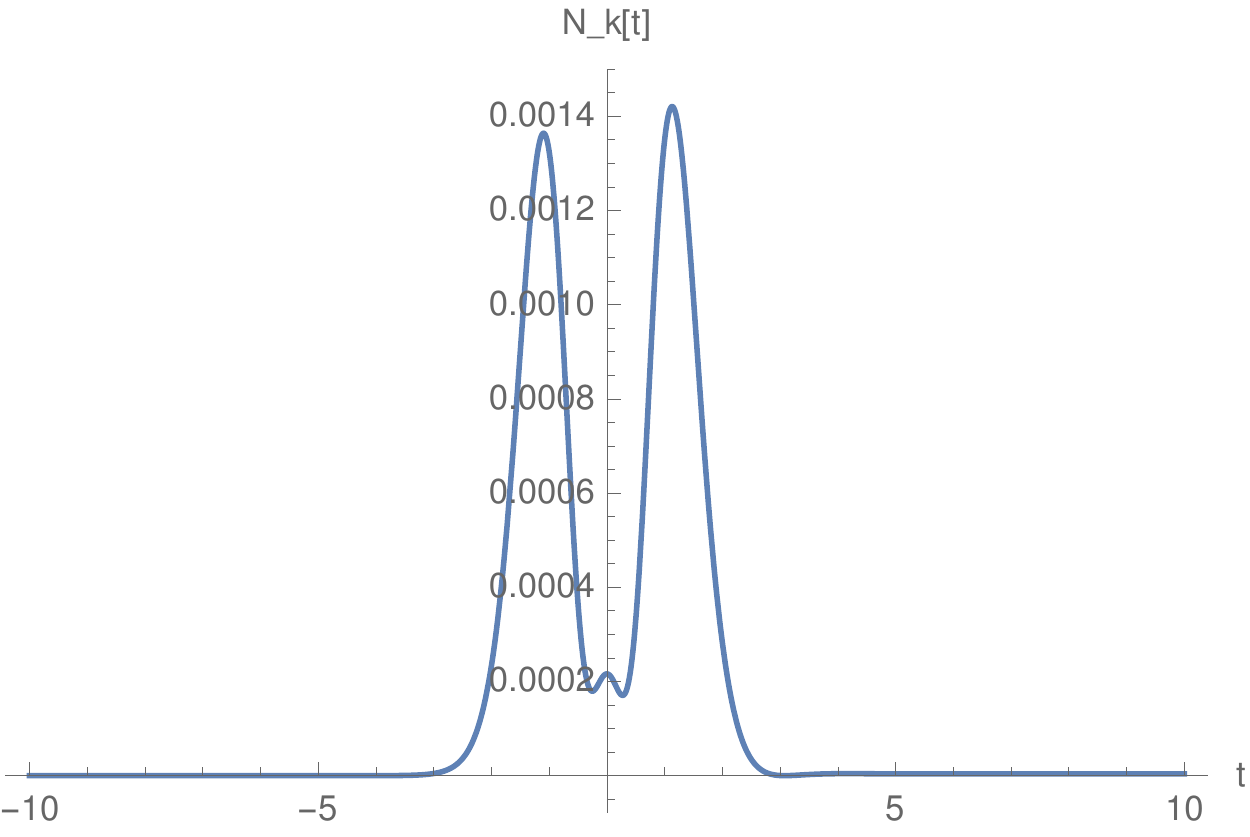}
\end{center}
\caption{Fermionic ${\cal N}_{\bf\tilde  k}(t)$ for the $p= \half \bigl(\pm \sqrt{21} -1\bigr)$ soliton.} 
\label{fig-m4}
\end{figure}

A numerical analysis using the Quantum Vlasov Equation \eqref{qvespin} suggests that
the above solitonic fields \eqref{solitons}
are non-pair creating also in the fermionic case, 
although  for different values of the parameter $ p$:
\begin{eqnarray}
p = \half\bigl(\pm \sqrt{m^2+m+1}-1\bigr), \quad
m=2,4,6,\ldots
\label{magic}
\end{eqnarray}
As an example, let us show the result of such a numerical evaluation for the simplest cases
$ m=2, p = \half \bigl(\pm \sqrt{7} -1\bigr)$ (Fig. \ref{fig-m2})
and $ m=4, p = \half \bigl(\pm \sqrt{21} -1\bigr)$ (Fig. \ref{fig-m4}).

\vspace{100pt}

This should be compared to the case $ p=1$, which would not pair create in the
scalar case but clearly does so here (Fig. \ref{fig-p1ferm}):

\begin{figure}[htp]
\vspace{-10pt}
\begin{center}
\includegraphics[scale=.45]{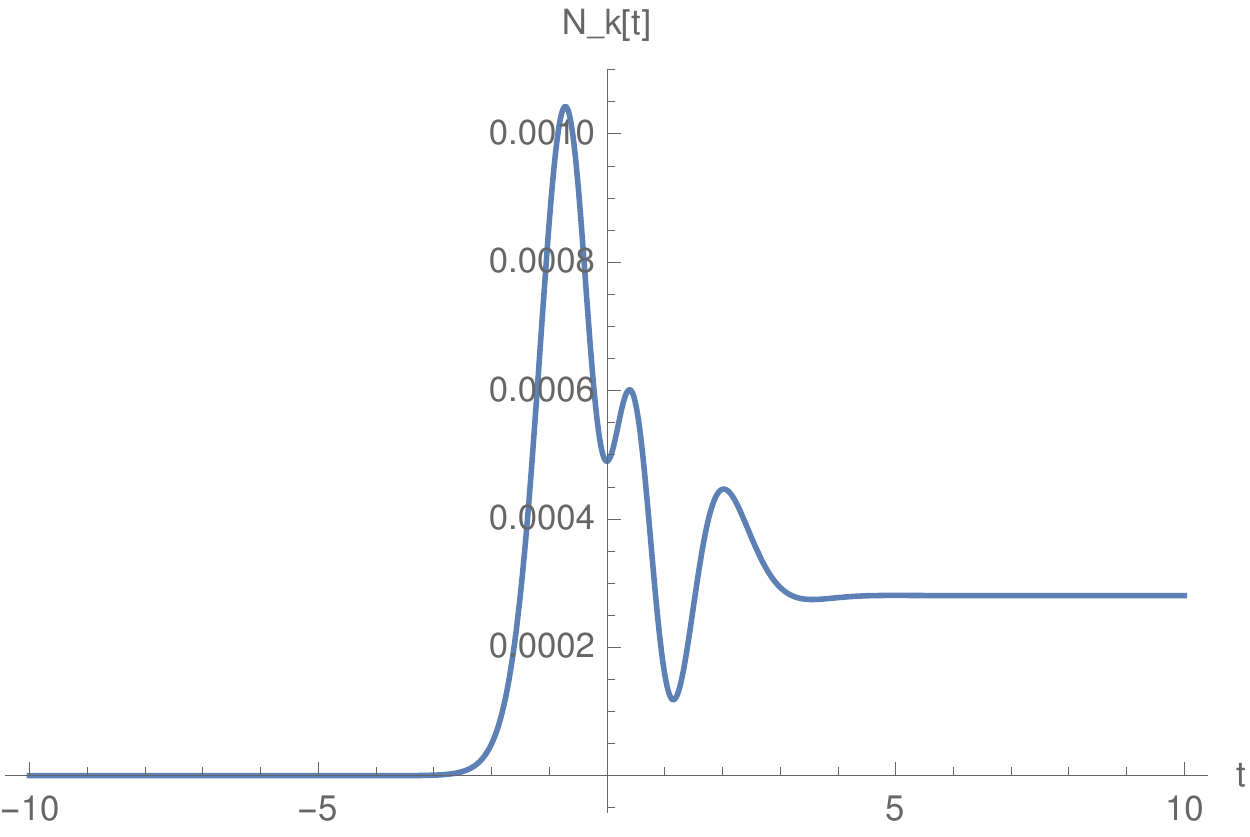}
\end{center}
\caption{Fermionic ${\cal N}_{\bf\tilde  k}(t)$ for the $p=1$ soliton.} 
\label{fig-p1ferm}
\end{figure}

\section{Summary and Outlook}

To summarize,

\begin{itemize}

\item
We have presented an evolution equation for fermions in a purely time-dependent
electric field that can be seen as a fermionic generalization of the Gelfand-Dikii equation.

\item
Solitonic electric fields allow one to tune the pair creation rate to disappear at some prescribed
longitudinal momentum for both scalars and spinors, although not for the same field.

\item
Thus they provide an example of  spin selection  in the pair creation process, and also a step towards 
``designer fields'', that is, the construction of fields with a prescribed pair-creation spectrum \cite{hebenstreit}. 

\item
It remains to find an analytic derivation of the ``magic numbers'' \eqref{magic}.

\end{itemize}

A more extensive study of fermionic pair creation in solitonic fields is in progress \cite{wip}. 

\ack
The work of SPK was supported by the Institute for Basic Science (IBS) under IBS-R012-D1.
CS thanks the CIC of UMSNH for financial support.


\section*{References}
\begin{thebibliography}{9}

\bibitem{dabdun}
R Dabrowski and G V Dunne, Time dependence of adiabatic particle number, Phys. Rev.
D 94, 065005 (2016), arXiv:1606.00902 [hep-th].

\bibitem{84}
S P Kim and C Schubert, 
Non-adiabatic Quantum Vlasov Equation for Schwinger Pair Production,
Phys .Rev. D {\bf 84} 125028 (2011), arXiv:1110.0900 [hep-th].

\bibitem{kescm91}
Y Kluger, J~M Eisenberg, B Svetitsky, F Cooper and E Mottola, 
Pair production in a strong electric field,
Phys. Rev. Lett. {\bf 67} (1991) 2427. 

\bibitem{kescm92}
Y Kluger, J~M Eisenberg, B Svetitsky, F Cooper and E Mottola,
Fermion pair production in a strong electric field,
Phys. Rev. D {\bf 45} (1992) 4659.

\bibitem{sbrspt}
S Schmidt, D Blaschke, G R\"opke, S~A Smolyansky,  A~V Prozorkevich and V~D Toneev,
A quantum kinetic equation for particle production in the Schwinger mechanism,
Int. J. Mod. Phys. E {\bf 7} (1998) 709, arXiv: hep-ph/9809227.

\bibitem{healgi}
F~Hebenstreit, R~Alkofer and H~Gies, 
Pair production beyond the Schwinger formula in time-dependent electric fields,
Phys.\ Rev.\ D {\bf 78}, 061701 (2008), arXiv:0807.2785.

\bibitem{geldik} 
I~M~Gelfand and ~A~Dikii,
Asymptotic Behavior of the Resolvent of Sturm-Liouville Equations and the Algebra of the Korteweg-De Vries Equations, Russ.\ Math.\ Surveys {\bf 30}, 5 (1975).

\bibitem{101}
 A Huet, S P Kim and C Schubert, 
 Vlasov equation for Schwinger pair production in a time-dependent electric field,
 Phys. Rev. D {\bf 90} (2014) 125033, arXiv:1411.3074 [hep-th].

\bibitem{basefr}
A B Balantekin, J E Seger and S H Fricke,
Dynamical effects in pair production by electric fields,
Int. J. Mod. Phys. A {\bf 6} (1991) 695. 

\bibitem{fgks}
A M Fedotov, E G Gelfer, K Yu Korolev and S A Smolyansky, 
Kinetic equation approach to pair production by a time-dependent electric field,
{\it Phys. Rev. D}, {\bf 83}, 025011 (2011), arXiv:1008.2098 [hep-ph].

\bibitem{hebenstreit}
F~Hebenstreit,
The inverse problem for Schwinger pair production,
Phys. Lett. B {\bf 753} (2016) 336, arXiv:1509.08693 [hep-ph].

\bibitem{wip}
N Ahmadiniaz,  A M Fedotov, E G Gelfer, S P Kim and C Schubert, 
Generalized Gelfand-Dikii equation and solitonic electric fields for fermionic Schwinger pair production,
in preparation. 

\end{thebibliography}
\end{document}